\documentclass{article}
\usepackage{amssymb}
\usepackage{amsmath}

\input{tcilatex}
\begin{document}

\title{ Influence of initial correlations on evolution over time of an open\\
quantum system \ \ \ }
\author{Victor F. Los \and Institute of Magnetism, Nat. Acad. Sci. and Min.
Edu. Sci. \and of Ukraine, 36-b Vernadsky Blvd., 03142, Kiev, Ukraine }

\begin{abstract}
A novel approach to accounting for the influence of initial system-bath
correlations on the dynamics of an open quantum system, based on the
conventional projection operator technique, is suggested. To avoid the
difficulties of treating the initial correlations, the conventional
Nakajima-Zwanzig inhomogeneous generalized master equations (GMEs) for a
system's reduced statistical operator and correlation function are exactly
converted \ into the homogeneous GMEs (HGMEs), which take into account the
initial correlations in the kernel governing the evolution of these HGMEs.
In the second order (Born) approximation in the system-bath interaction, the
obtained HGMEs are local in time and valid at all timescales. They are
further specialized for a realistic equilibrium Gibbs initial (at $t=t_{0}$)
system+bath state (for a system reduced statistical operator an external
force at $t>t_{0}$ is applied) and then for a bath of oscillators (Boson
field). As an example, the evolution of a selected quantum oscillator (a
localized mode) interacting with a Boson field (Fano-like model) is
considered at different timescales. It is shown explicitly how the initial
correlations influence the oscillator evolution process. In particular, it
is shown that the equilibrium system's correlation function acquires at the
large timescale the additional constant phase factor conditioned by survived
initial system-bath correlations.
\end{abstract}

\maketitle

\section{Introduction}

Rigorous derivation of the reliable, tractable and effective evolution
equations for the measurable values (statistical expectations)
characterizing a nonequilibrium state of a many-particle system remains a
principal task of statistical physics. The derivation is conventionally
started from the Liouville-von Neumann equation for a statistical operator
(distribution function) of the whole system, which depends on the enormous
number of variables and thus is practically useless. Fortunately, for
practical purposes, one needs the equations only for reduced statistical
operators (distribution functions) for a system of interest depending on a
much smaller number of variables and which can be obtained by integration
off the environment (irrelevant) variables. It is expected that these
equations, although obtained from underlying reversible microscopic
many-particle dynamics (Liouville-von Neumann equation), should generally be
the equations converting into irreversible kinetic ones on some timescale.

In order to derive the equation for reduced statistical operator, several
assumptions are usually made. One of them is related to the correlations
between a selected system under consideration and its environment in the
initial state of a full system (initial correlations). For example, all
derivations of the Boltzmann equation for a many-particle system or the
Lindblad equation for evolution of a system interacting with a bath are
mainly based on either factorizing-type initial conditions (random phase
approximation (RPA) or "molecular chaos") corresponding to uncorrelated
initial state or on Bogoliubov's principle of weakening of initial
correlations \cite{Bogoliubov}. The uncorrelated initial state is not very
realistic \cite{van Kampen} and the Bogoliubov principle is not always
applicable, e.g., when the correlations do not damp with time. Such a
situation is realized, e.g., in a quantum many-body system, where the
quantum correlations, caused by the particles' statistics, do not damp with
time \cite{Los JSTAT 2024}, or for an equilibrium initial state of the whole
system \cite{van Kampen,Los JSTAT 2022}. More generally, the disregarding of
the initial (at the initial time moment $t_{0}$) correlation (irrelevant)
term, e.g., in the inhomogeneous Nakajima-Zwanzig generalized master
equation (GME) \cite{Nakajima,Zwanzig} for the relevant (reduced) part of a
statistical operator of interest, implies, in fact, the "propagation of
chaos", i.e. the absence of correlations also at $t>t_{0}$ (see, e.g., \cite%
{van Kampen,Wallace}). The latter has not been proved yet \cite{Kac}. Since
there are no convincing arguments for neglecting the initial correlations 
\cite{van Kampen,Wallace}, it would be desirable to effectively include
initial correlations into consideration and thus to have a completely closed
(homogeneous) and valid on all timescales evolution equation for a reduced
statistical operator. The approaches, like the Bogoliubov principle of
weakening of initial correlations, result in the closed homogeneous
equations for a system of interest describing the evolution on the large
enough timescale, when correlations might damp, and, therefore, are not able
to describe the influence of initial correlations on the entire evolution
process. The effective accounting in the evolution equation for initial
correlations assumes that such an equation at the large timescale would not
lead to the appearance of the so called "secular" terms growing with time
as, e.g., it is the case when one is trying to solve the inhomogeneous
equations of the BBGKY (after the names of Bogoliubov, Born, Green, Kirkwood
and Yvon) chain\ \cite{Bogoliubov}. Generally, the homogeneous equations
have a much wider range of applicability with regard to the timescale than
the inhomogeneous ones.

Of a special interest is the situation when a system interacts with a
stationary environment (a bath), i.e., an open quantum system. Open quantum
systems have become an active area of research, owing to its potential
applications in many different fields such as quantum gases, quantum optics,
quantum information processing, quantum computing (to mention a few), and
the general problems of statistical physics \cite{Breuer}. In this area, the
evolution of a system is conventionally described by the Lindblad equation
(see, \cite{Lindblad}, \cite{Kossakowski et al} and \cite{Breuer}), which is
considered now a cornerstone of the theory of open quantum systems.
Actually, this equation follows from the Redfield equation \cite{Redfield
1957,Blum 1981}, which is a local in time equation for a system $S$ density
matrix $\rho _{S}(t)$ obtained from the von-Neumann equation of motion for
the combined (system+environment) statistical operator in the second (Born)
approximation in a weak system-environment interaction with additional
assumptions that the total system-bath density matrix $\rho (t)$ can be
factorized for any time $t$, i.e., $\rho (t)=$ $\rho _{S}(t)\rho _{B}$,
where $\rho _{B}$ is a bath density matrix, and that the system density
matrix at any retarded time $\rho _{S}(t^{\prime })$ ($t^{\prime }<t$) can
be replaced by that at the present time $\rho _{S}(t)$ (Markovian
approximation). To make the Redfield master equation fully Markovian one,
the Born-Markov approximation is further applied, which is justified when
the bath correlation time $\tau _{B}$ is small compared with the relaxation
time of a system $\tau _{S}$, i.e.,$\tau _{B}\ll \tau _{S}$. The Born-Markov
approximation means that we consider the system evolution at a large
timescale $t\thicksim \tau _{S}\gg \tau _{B}$. Generally, the Markovian
Redfield equation does not guarantee the dynamics complete positivity, and
in order to achieve this goal, the additional approximation, which involves
an averaging over the rapidly oscillating terms in the master equation (the
rotating wave approximation), is needed. This procedure, which eliminates
the very rapidly oscillating during the time $\tau _{S}$ terms, means that
the evolution equation is valid for a large timescale $t\thicksim \tau _{S}$%
. The final quantum master equation, which results from the mentioned
approximations, is the Lindblad equation preserving all properties of the
density matrix (including complete positivity). Therefore, the Lindblad
equation (like the Redfield equation) has its fundamental limitations (see
also \cite{Phys.Rev.A105032208(2022)}).

The essential conventional assumption, made at the derivation of the
Redfield (and the Lindblad) equation, is that the system-environment
correlations (including the correlations at the initial moment of time $%
t_{0} $) are ignored. There are several approaches for including the
correlations either for specific Hamiltonians (see, e.g., \cite%
{Phys.Rev.A83032102(2011)}), or in the frameworks of completely positive
trace preserving (CPTP) maps \cite{Phys.Rev.A100042120(2019)} and of the
Lindblad-like equations (\cite{Phys.Rev.X(2020)}).

In this paper, we suggest the different approach to the initial correlation
problem based on the modified standard time-independent projection operator
formalism allowing for exact transformation of the conventional
Nakajima-Zwanzig-like inhomogeneous generalized master equations (GMEs) for
the relevant parts of a system statistical operator or correlation function,
containing initial correlations in the inhomogeneous (irrelevant) initial
condition term (a source), into the homogeneous generalized master equations
(HGMEs) with no irrelevant source term. These exact HGMEs contain initial
correlations in the modified flow and collision terms. The parameter of
initial correlations is defined by a series in the product of irrelevant
parts of statistical operator or correlation function and the inverse of the
corresponding relevant parts. In the second order approximation in the
system-environment interaction, the obtained HGMEs are local in time and
contain general initial system-bath interaction in the kernel governing the
evolution of these equations. They exactly describe in the Born
approximation the system evolution on any timescale (no Bogoliubov's
principle of weakening of initial correlations or the Markov approximation
is applied) and have an additional terms defined by initial correlations.
These terms are considered for an interesting (and quite realistic) case of
the system+bath equilibrium Gibbs initial state, when initial correlations
do not damp with time. The equations for $\rho _{S}(t)$ (at $t>t_{0}$, when
an external force is applied) and system's correlation function are further
specialized for the bath of oscillators (Boson field) and then for a quantum
oscillator as a system of interest (Fano-like system). These equations are
valid on any timescale and describe the evolution of a localized Bose mode.
The solutions of equations show that the equilibrium initial correlations
influence the selected system's evolution process. For short times, $%
\overline{\omega }t\ll 1$ ($\overline{\omega }$ is a bath characteristic
frequency), the relaxation process is quadratic in time, and for long times, 
$\overline{\omega }t\gg 1$, the evolution exhibits the standard behavior,
but with additional phase shift for the equilibrium system's correlation
function conditioned by survived equilibrium initial correlations.
Consequences of such a behavioral for the observables are discussed.

\section{Time-independent projection operator formalism with no "molecular
chaos"-type approximation}

\subsection{Homogeneous equation for the relevant statistical operator}

We start with the von-Neumann equation for a statistical operator $F(t)$ of $%
N$ ($N\gg 1$) quantum particles 
\begin{equation}
\frac{\partial }{\partial t}F(t)=L(t)F(t).  \label{1}
\end{equation}%
Here, $F(t)$ satisfies the normalization condition 
\begin{equation}
TrF(t)=1,  \label{1'}
\end{equation}%
and $L$ is the Liouville superoperator acting on an arbitrary operator $A(t)$
as 
\begin{equation}
L(t)A(t)=\frac{1}{i\hbar }[H(t),A(t)],\exp [L(t)]A(t)=\exp [H(t)/i\hbar
]A(t)\exp [-H(t)/i\hbar ],  \label{1''}
\end{equation}%
where $[,]$ is a commutator and $H(t)$ is the system's Hamiltonian,
generally dependent on time.

The formal solution to Eq. (\ref{1}) is%
\begin{equation}
F(t)=U(t,t_{0})F(t_{0}),U(t,t_{0})=T\exp \left[ \tint%
\limits_{t_{0}}^{t}dsL(s)\right] ,  \label{1'''}
\end{equation}%
where $T$ denotes the chronological time-ordering operator, which orders the
product of time-dependent operators such that their time-arguments increase
from right to left, and $F(t_{0})$ is the statistical operator at some
initial moment of time $t_{0}$ (initial condition).

We employ first the standard projection operator technique \cite{Nakajima}, 
\cite{Zwanzig}, \cite{Prigogine (1962)}. By applying the time-independent
projection operators $P=P^{2}$ and $Q=Q^{2}=1-P$ ($QP=PQ=0$) to Eq. (\ref{1}%
), it is easy to obtain the equations for the relevant $f_{r}(t)=PF(t)$ and
irrelevant $f_{i\text{ }}(t)=QF(t)$ parts of $F(t)$ 
\begin{align}
\frac{\partial }{\partial t}f_{r}(t)& =PL(t)[f_{r}(t)+f_{i}(t)],  \notag \\
\frac{\partial }{\partial t}f_{i}(t)& =QL(t)[f_{r}(t)+f_{i}(t)].  \label{3}
\end{align}%
A relevant part of $F(t)$ is defined in a way permitting the calculation of
the average values (observables) of the operators depending on the much
smaller number $s\ll N$ of relevant variables than that of the total system.
This is typically achieved by selecting a projector $P$ which integrates off
the irrelevant variables excessive for the calculation of observables of
interest.

A formal solution to the second Eq. (\ref{3}) has the form%
\begin{equation}
f_{i}(t)=\tint\limits_{t_{0}}^{t}dt^{\prime }S(t,t^{\prime })QL(t^{\prime
})f_{r}(t^{\prime })+S(t,t_{0})f_{i}(t_{0}),  \label{3a}
\end{equation}%
where 
\begin{equation}
S(t,t_{0})=T\exp \left[ \tint\limits_{t_{0}}^{t}dsQL(s)\right] .  \label{3b}
\end{equation}%
Inserting this solution into the first Eq. (\ref{3}), we obtain the
conventional exact time-convolution generalized master equation (TC-GME)
known as the Nakajima-Zwanzig equation for the relevant part of the
statistical operator (see also \cite{Breuer})%
\begin{equation}
\frac{\partial f_{r}(t)}{\partial t}=PL(t)[f_{r}(t)+\tint%
\limits_{t_{0}}^{t}dt^{\prime }S(t,t^{\prime })QL(t^{\prime
})f_{r}(t^{\prime })+S(t,t_{0})f_{i}(t_{0})].  \label{5}
\end{equation}

Serving as a basis for many applications, Eq. (\ref{5}), nevertheless,
contains the undesirable and in general non-negligible inhomogeneous term
(the last term in the right hand side of (\ref{5})), which depends (via $f_{i%
\text{ }}(t_{0})$) on the same large number of variables as the statistical
operator $F(t_{0})$ at the initial instant $t_{0}$. Therefore, Eq. (\ref{5})
does not provide for a complete reduced description of a multiparticle
system in terms of relevant (reduced) statistical operator. Applying
Bogoliubov's principle of weakening of initial correlations \cite{Bogoliubov}
allowing to eliminate the influence of $f_{i}(t_{0})$ on the large enough
timescale $t-t_{0}\gg t_{cor}$ ($t_{cor}$ is the correlation time caused by
inter-particle interaction) or using a factorized initial condition,when $%
f_{i\text{ }}(t_{0})=QF(t_{0})=0$, one can achieve the above-mentioned goal
and obtain the homogeneous GME for $f_{r}(t)$, i.e. Eq. (\ref{5}) with no
initial condition term. However, obtained in such a way homogeneous GME is
either approximate and valid only on a large enough time scale (when all
initial correlations vanish) or applicable only for a rather artificial
(actually not realistic), as pointed in \cite{van Kampen}, initial
conditions (no correlations at an initial instant of time). As to our
understanding, there is no satisfactory way for eliminating the irrelevant
initial condition term (see, e.g., \cite{Wallace}).

In order to obtain an exact homogeneous equation for the relevant part of a
statistical operator, one can try to transfer the inhomogeneous initial
correlations term in the right hand side of Eq. (\ref{5}) to the
(super)operator acting on the relevant part $f_{r}(t)$. To achieve this
goal, we suggest to present the initial (irrelevant) term $%
f_{i}(t_{0})=QF(t_{0})$ as a following exact identity%
\begin{align}
f_{i}(t_{0})&
=f_{i}(t_{0})F^{-1}(t_{0})U^{-1}(t,t_{0})(P+Q)U(t,t_{0})F(t_{0})  \notag \\
& =K_{0}U^{-1}(t,t_{0})[f_{r}(t)+f_{i}(t)],  \notag \\
K_{0}& =f_{i}(t_{0})F^{-1}(t_{0}),U^{-1}(t,t_{0})=T_{-}\exp
[-\tint\limits_{t_{0}}^{t}dsL(s)],  \label{6}
\end{align}%
where $U^{-1}(t,t_{0})$ is the backward-in-time evolution operator, $T_{-}$
is the antichronological time-ordering operator arranging the time-dependent
operators $L(s)$ in such a way that the time arguments increase from left to
right, $U^{-1}(t,t_{0})U(t,t_{0})=1$, $F^{-1}(t_{0})$ is inverse of $%
F(t_{0}) $, $F^{-1}(t_{0})F(t_{0})=1$, and $P+Q=1$. Hence, additional
identity (\ref{6}) is obtained by multiplying the irrelevant part by unity $%
F^{-1}(t_{0})F(t_{0})$ and inserting $U^{-1}(t,t_{0})U(t,t_{0})=1$ and $%
P(t)+Q(t)=1$.

In (\ref{6}), we introduce the parameter of initial correlations 
\begin{align}
K_{0}&
=f_{i}(t_{0})F^{-1}(t_{0})=f_{i}(t_{0})[f_{r}(t_{0})+f_{i}(t_{0})]^{-1} 
\notag \\
& =f_{i}(t_{0})f_{r}^{-1}(t_{0})[1+f_{i}(t_{0})f_{r}^{-1}(t_{0})]^{-1} 
\notag \\
& =(1-K_{0})f_{i}(t_{0})f_{r}^{-1}(t_{0}),  \label{7}
\end{align}%
which is a series in $f_{i}(t_{0})f_{r}^{-1}(t_{0})$.

We now have two equations, (\ref{3a}) and (\ref{6}), relating $f_{i}(t)$ to $%
f_{i}(t_{0})$. Finding $f_{i}(t_{0})$ from these equations as a function of $%
f_{r}(t)$ and substituting it in (\ref{5}), we obtain the equation

\begin{equation}
\frac{\partial f_{r}(t)}{\partial t}=PL(t)R(t,t_{0})[f_{r}(t)+\tint%
\limits_{t_{0}}^{t}dt^{\prime }S(t,t^{\prime })QL(t^{\prime
})f_{r}(t^{\prime })],  \label{8}
\end{equation}%
where operator $R(t,t_{0})$ is defined as%
\begin{align}
R(t,t_{0})& =1+K(t,t_{0}),  \notag \\
K(t,t_{0})& =S(t,t_{0})\left[ 1-K_{0}(t,t_{0})\right]
^{-1}K_{0}U^{-1}(t,t_{0}),  \notag \\
K_{0}(t,t_{0})& =K_{0}U^{-1}(t,t_{0})S(t,t_{0}).  \label{9}
\end{align}

Thus, we have obtained the exact integro-differential Eq. (\ref{8}), which
is the homogeneous one (in contrast to Eq. (\ref{5})) and takes into account
the initial correlations and their dynamics on an equal footing with
collisions via the modification of the (super)operator (memory kernel) of
GME (\ref{5}) acting on the relevant part of the statistical operator $%
f_{r}(t)$. The obtained exact kernel of Eq. (\ref{8}) can serve as a
starting point for consecutive perturbation expansions. In many cases such
expansions of the homogeneous equations (like (\ref{8})) have much broader
range of validity than that for the inhomogeneous ones (like (\ref{5})).

\subsection{Homogeneous equation for system's correlation function}

Another useful approach to studying the reduced dynamics of a many-particle
system is the correlation functions approach. We divide the total system of $%
N$ particle into a system of interest $S$ with a number of particles $s\ll N$
and its environment (the rest of particles). Let us consider the following
two-time correlation function for the time-dependent operators $\alpha
_{S}(t)=e^{-Lt}\alpha _{S}(0)$ and $\beta _{S}(t)=e^{-Lt}\beta _{S}(0)$
related to the system $S$ 
\begin{eqnarray}
\langle \alpha _{S}(0)\beta _{S}(t)\rangle &=&Tr_{S}[\beta _{S}(0)\rho
_{S}^{\alpha }(t)],\rho _{S}^{\alpha }(t)=Tr_{B}F_{\alpha }(t)],  \notag \\
F_{\alpha }(t) &=&e^{Lt}F_{\alpha }(0),F_{\alpha }(0)=\rho _{tot}(0)\alpha
_{S}(0),  \label{9a}
\end{eqnarray}%
where the superoperator $L$ is given by (\ref{1''}) (with time-independent $%
L $), and the averaging symbol $\langle ...\rangle =Tr_{S+B}[...\rho
_{tot}(0)] $ is defined as the trace $Tr_{S+B}$ over the total $S+B$ system
degrees of freedom with the time-independent statistical operator $\rho
_{tot}(0)=\rho _{tot}$ at the initial time moment $t_{0}=0$ (the Heisenberg
picture), which generally is not a statistical operator for $S+B$ system in
an equilibrium state. We also used the invariance of trace under the cyclic
permutation of operators. The superoperator $F_{\alpha }(t)$ is subject to
the following equation%
\begin{equation}
\frac{\partial }{\partial t}F_{\alpha }(t)=LF_{\alpha }(t),  \label{9b}
\end{equation}%
which defines the evolution of the correlation function (\ref{9a}) in time .

As earlier, with the help of the projection operators $P$ and $Q=1-P$ and
taking into account that in Eq. (\ref{9b}) $L$ does not depend on time, we
can write down the Nakajima-Zwanzig equation (\ref{5}) for the relevant part
of the superoperator $F_{\alpha }(t)$, $f_{r}^{\alpha }(t)=PF_{\alpha }(t)$,
as 
\begin{equation}
\frac{\partial f_{r}^{\alpha }(t)}{\partial t}=PL[f_{r}^{\alpha
}(t)+\tint\limits_{0}^{t}dt^{\prime }e^{QL(t-t^{\prime })}QLf_{r}^{\alpha
}(t^{\prime })+e^{QLt}f_{i}^{\alpha }(0)],  \label{9c}
\end{equation}%
where the irrelevant part of $F_{\alpha }(t)$ is $f_{i}^{\alpha
}(t)=QF_{\alpha }(t)$. As it is seen from Eq. (\ref{9c}), the projector $P$
selects the relevant part of $F_{\alpha }(t)$, which is supposed to be
sufficient for obtaining the closed equation for the evolution of
correlation function for system's $S$ operators, i.e., an operator $P$
projects the large-dimensional Hilbert space of the total system on the
smaller dimensional space of the system of interest $S$. But Eq. (\ref{9c})
is the inhomogeneous one which contains the irrelevant initial condition
term $f_{i}^{\alpha }(0)=(\rho _{tot}-P\rho _{tot})\alpha _{S}(0)$.

To make this equation homogeneous (completely closed), the above described
procedure can be employed. As a result, we arrive at the following
homogeneous equation

\begin{equation}
\frac{\partial f_{r}^{\alpha }(t)}{\partial t}=PLR^{\alpha
}(t)[f_{r}^{\alpha }(t)+\tint\limits_{0}^{t}dt^{\prime }e^{QL(t-t^{\prime
})}QLf_{r}^{\alpha }(t^{\prime })],  \label{9d}
\end{equation}%
where operator $R^{\alpha }(t)$ is defined as%
\begin{align}
R^{\alpha }(t)& =1+K^{\alpha }(t),  \notag \\
K^{\alpha }(t)& =e^{QLt}\left[ 1-K_{0}^{\alpha }(t)\right]
^{-1}K_{0}^{\alpha }e^{-Lt},  \notag \\
K_{0}^{\alpha }(t)& =K_{0}^{\alpha }e^{-Lt}e^{QLt},  \notag \\
K_{0}^{\alpha }& =f_{i}^{\alpha }(0)F_{\alpha }^{-1}(0)=(1-K_{0}^{\alpha
})f_{i}^{\alpha }(0)[f_{r}^{\alpha }(0)]^{-1}.  \label{9e}
\end{align}

\section{System in a bath}

For the case under consideration, when a system of interest $S$ interacts
with another quantum stationary system $B$, called bath, the Hamiltonian of
the whole $S+B$ system can be presented as 
\begin{equation}
H(t)=H_{S}(t)+H_{B}+H_{SB},  \label{10}
\end{equation}%
where $H_{S}(t)$ is related to a system $S$ and can depend on time through
an applied external force, $H_{B}$ and $H_{SB}$ are the Hamiltonians of a
bath and of a system-bath interaction, respectively. The corresponding
Liouville superoperator is%
\begin{equation}
L(t)=L_{S}(t)+L_{B}+L_{SB}.  \label{11}
\end{equation}

We are interested in finding the evolution equation for statistical operator
of the system $S$ 
\begin{equation}
\rho _{S}(t)=Tr_{B}F(t)  \label{12}
\end{equation}%
from Eq. (\ref{8}) ($Tr_{B}$ is the partial trace of the bath degrees of
freedom). To this end, it is convenient to introduce the following
projection operators%
\begin{equation}
P=\rho _{B}Tr_{B},Q=1-P,Tr_{B}\rho _{B}=1,  \label{13}
\end{equation}%
where $\rho _{B}$ stands for the normalized statistical operator of a bath.
Then the relevant and irrelevant parts of the statistical operator $F(t)$
are 
\begin{equation}
f_{r}(t)=PF(t)=\rho _{B}\rho _{S}(t),f_{i}(t)=QF(t)=F(t)-\rho _{B}\rho
_{S}(t).  \label{14}
\end{equation}

Then, Eq. (\ref{8}) can be rewritten in a more specific and simple form for
the Liouvillian (\ref{11}) and projection operators (\ref{13}) if we take
into consideration the following properties 
\begin{equation}
PL_{S}(t)Q=QL_{S}(t)P=0,PL_{B}=L_{B}P=0,PL_{B}Q=QL_{B}P=0.  \label{16}
\end{equation}%
where we took into account that $[H_{B},\rho _{B}]=0$. We also assume that%
\begin{equation}
\langle H_{SB}\rangle _{B}=Tr_{B}H_{SB}\rho _{B}=0  \label{16a}
\end{equation}%
and, therefore, 
\begin{equation}
PL_{SB}P=0,  \label{17}
\end{equation}%
which is a rather typical situation (see below).

Now, Eq. (\ref{8}) can be rewritten as 
\begin{eqnarray}
\frac{\partial f_{r}(t)}{\partial t} &=&L_{S}(t)f_{r}(t)+PL_{SB}\overline{K}%
(t,t_{0})f_{r}(t)  \notag \\
&&+PL_{SB}[1+\overline{K}(t,t_{0})]\tint\limits_{t_{0}}^{t}dt^{\prime }%
\overline{S}(t,t^{\prime })L_{SB}f_{r}(t^{\prime }),  \label{18}
\end{eqnarray}%
where $\overline{K}(t,t_{0})$ is defined by (\ref{9}) with $S(t,t^{\prime })$
replaced by 
\begin{eqnarray}
\overline{S}(t,t^{\prime }) &=&T\exp \left\{ \tint\limits_{t^{\prime
}}^{t}ds[L_{0}(s)+QL_{SB}]\right\} ,  \notag \\
L_{0}(s) &=&L_{S}(s)+L_{B}  \label{19}
\end{eqnarray}

Taking $Tr_{B}$ from both sides of Eq. (\ref{18}) and using definitions (\ref%
{13}) and (\ref{14}), we obtain the following exact equation for a system's
statistical operator%
\begin{eqnarray}
\frac{\partial \rho _{S}(t)}{\partial t} &=&L_{S}(t)\rho _{S}(t)+Tr_{B}L_{SB}%
\overline{K}(t,t_{0})\rho _{B}\rho _{S}(t)  \notag \\
&&+Tr_{B}L_{SB}[1+\overline{K}(t,t_{0})]\tint\limits_{t_{0}}^{t}dt^{\prime }%
\overline{S}(t,t^{\prime })L_{SB}\rho _{B}\rho _{S}(t^{\prime }).  \label{20}
\end{eqnarray}

This equation differs from the standard form of such an equation (see, e.g., 
\cite{Breuer}) by the additional terms in the kernel containing $\overline{K}%
(t,t_{0})$, which account for the influence of initial system-bath
correlations on the evolution of the system. We also note, that
conventionally used "molecular chaos"-type approximation, $f_{r}(t^{\prime
})=\rho _{B}\rho _{S}(t^{\prime })$ (including the initial time moment $%
t^{\prime }=t_{0}$), is not used when going from Eq. (\ref{18}) to Eq. (\ref%
{20}).

In the correlation function approach, when we have a time-independent $L_{S}$
($H_{S}$), the relevant and irrelevant parts of $F_{\alpha }(t)$ selected by
the projector (\ref{13}) are%
\begin{eqnarray}
f_{r}^{\alpha }(t) &=&PF_{\alpha }(t)=\rho _{B}\rho _{S}^{\alpha
}(t),f_{i}^{\alpha }(t)=QF_{\alpha }(t)=F_{\alpha }(t)-\rho _{B}\rho
_{S}^{\alpha }(t),  \notag \\
\rho _{S}^{\alpha }(t) &=&Tr_{B}F_{\alpha }(t).  \label{20'}
\end{eqnarray}

Then, using (\ref{16}) and (\ref{17}), Eq. (\ref{9d}) can be simplified as%
\begin{eqnarray}
\frac{\partial f_{r}^{\alpha }(t)}{\partial t} &=&L_{S}f_{r}^{\alpha
}(t)+PL_{SB}\overline{K}^{\alpha }(t)f_{r}^{\alpha }(t)  \notag \\
&&+PL_{SB}[1+\overline{K}^{\alpha }(t)]\tint\limits_{0}^{t}dt^{\prime
}e^{(L_{0}+QL_{SB})(t-t^{\prime })}L_{SB}f_{r}^{\alpha }(t^{\prime }),
\label{20''}
\end{eqnarray}%
where $\overline{K}^{\alpha }(t)$ is defined by (\ref{9e}) with the
substitution $e^{QLt}\rightarrow \exp (L_{0}+QL_{SB}).$We remind, that for
the correlation function case, the Hamiltonian is time-independent and $%
L_{0}=L_{S}+L_{B}$.

Again, by applying $Tr_{B}$ to Eq. (\ref{20''}) from the left, we obtain the
following exact equation for $\rho _{S}^{\alpha }(t)$%
\begin{eqnarray}
\frac{\partial \rho _{S}^{\alpha }(t)}{\partial t} &=&L_{S}\rho _{S}^{\alpha
}(t)+Tr_{B}L_{SB}\overline{K}^{\alpha }(t)\rho _{B}\rho _{S}^{\alpha }(t) 
\notag \\
&&+Tr_{B}L_{SB}[1+\overline{K}^{\alpha }(t)]\tint\limits_{0}^{t}dt^{\prime
}e^{(L_{0}+QL_{SB})(t-t^{\prime })}L_{SB}\rho _{B}\rho _{S}^{\alpha
}(t^{\prime }),  \label{20'''}
\end{eqnarray}%
which defines the evolution of the correlation function (\ref{9a}) according
to%
\begin{equation}
\frac{\partial }{\partial t}\langle \alpha _{S}(0)\beta _{S}(t)\rangle
=Tr_{S}[\beta _{S}(0)\frac{\partial }{\partial t}\rho _{S}^{\alpha }(t)].
\label{20''''}
\end{equation}

Obtained Eqs. (\ref{20}) and (\ref{20'''}) are the homogeneous exact
evolution equations for a system's reduced statistical operator and
correlation function in the case of a system interacting with a bath which
exactly account for initial correlations.

\section{The Born approximation}

Equations (\ref{20}) and (\ref{20'''}) are exact but very complicated. Let
us consider the case of a weak system-bath interaction when $H_{SB\text{ }}$%
is proportional to some small parameter $\varepsilon $. The collision terms
(third in the r.h.s. of Eq.(\ref{20}) and Eq. (\ref{20'''})) are already
proportional to $\varepsilon ^{2}$. Then, we will restrict ourselves by the
second order in the system-bath interaction (the Born approximation). So, we
need to estimate (with regard to the power of $\varepsilon $) the initial
correlations parameters $\overline{K}(t,t_{0})$ and $\overline{K}^{\alpha
}(t)$. In the zero approximation in $\varepsilon $, 
\begin{eqnarray}
\overline{S}(t,t_{0}) &=&U_{0}(t,t_{0})=T\exp
[\tint\limits_{t_{0}}^{t}dsL_{0}(s)],  \notag \\
U_{0}^{-1}(t,t_{0}) &=&T_{-}\exp [-\tint\limits_{t_{0}}^{t}dsL_{0}(s)], 
\notag \\
\exp [(L_{0}+QL_{SB})t] &=&\exp (L_{0}t),e^{-Lt}=e^{-L_{0}t}.,  \label{20a}
\end{eqnarray}%
and, therefore, $\overline{K_{0}}(t,t_{0})=K_{0}$ and $\overline{K}%
_{0}^{\alpha }=K_{0}^{\alpha }$ (see (\ref{9}) and (\ref{9e})). Therefore, $%
\overline{K}(t,t_{0})=\overline{K}%
_{1}(t,t_{0})=U_{0}(t,t_{0})(1-K_{0})^{-1}K_{0}U_{0}^{-1}(t,t_{0})$, $%
\overline{K}^{\alpha }(t)=\overline{K}_{1}^{\alpha
}(t)=e^{L_{0}t}(1-K_{0}^{\alpha })^{-1}K_{0}^{\alpha }e^{-L_{0}t}$, and
because $(1-K_{0})^{-1}K_{0}=f_{i}(t_{0})f_{r}^{-1}(t_{0})$, $%
(1-K_{0}^{\alpha })^{-1}K_{0}^{\alpha }=f_{i}^{\alpha }(0)[f_{r}^{\alpha
}(0)]^{-1}$, as it follows from (\ref{7}) and (\ref{9e}), we finally have in
the adopted approximation%
\begin{eqnarray}
\overline{K}_{1}(t,t_{0})
&=&G_{SB}(t,t_{0})=U_{0}(t,t_{0})f_{i}(t_{0})f_{r}^{-1}(t_{0})U_{0}^{-1}(t,t_{0}),
\notag \\
\overline{K}_{1}^{\alpha }(t) &=&G_{SB}^{t}(t)=e^{L_{0}t}f_{i}^{\alpha
}(0)[f_{r}^{\alpha }(0)]^{-1}e^{-L_{0}t},  \label{21}
\end{eqnarray}%
where 
\begin{eqnarray}
f_{i}(t_{0})f_{r}^{-1}(t_{0}) &=&[F(t_{0})-\rho _{B}\rho _{S}(t_{0})][\rho
_{B}\rho _{S}(t_{0})]^{-1},\rho _{S}(t_{0})=Tr_{B}F(t_{0}),  \notag \\
f_{i}^{\alpha }(0)[f_{r}^{\alpha }(0)]^{-1} &=&(\rho _{tot}-\rho _{B}\rho
_{S}^{t})\alpha _{S}(0)[\rho _{B}\rho _{S}^{t}\alpha _{S}(0)]^{-1}  \notag \\
&=&(\rho _{tot}-\rho _{B}\rho _{S}^{t})[\rho _{B}\rho _{S}^{t}]^{-1},\rho
_{S}^{t}=Tr_{B}\rho _{tot}  \label{22}
\end{eqnarray}%
(see (\ref{14})) and (\ref{20'}). Note, that $G_{SB}^{t}(t)$ (\ref{21}) does
not depend on $\alpha _{S}(0)$ (see (\ref{22})). Estimating $[F(t_{0})-\rho
_{B}\rho _{S}(t_{0})]\thicksim \varepsilon $ and $(\rho _{tot}-\rho _{B}\rho
_{S}^{t})\thicksim \varepsilon $, we obtain that both functions (\ref{21}) $%
\thicksim \varepsilon $.

Thus, in the second order in $\varepsilon $, we have from Eqs. (\ref{20})
and (\ref{20'''}) the following equations for $\rho _{S}(t)$ and $\rho
_{S}^{\alpha }(t)$ after the change of integration variable $t-t^{\prime
}=\tau $%
\begin{eqnarray}
\frac{\partial \rho _{S}(t)}{\partial t} &=&L_{S}(t)\rho
_{S}(t)+Tr_{B}L_{SB}G_{SB}(t,t_{0})\rho _{B}\rho _{S}(t)  \notag \\
&&+Tr_{B}L_{SB}\tint\limits_{0}^{t-t_{0}}d\tau U_{0}(t,t-\tau )L_{SB}\rho
_{B}\rho _{S}(t-\tau ),  \notag \\
\frac{\partial \rho _{S}^{\alpha }(t)}{\partial t} &=&L_{S}\rho _{S}^{\alpha
}(t)+Tr_{B}L_{SB}G_{SB}^{t}(t)\rho _{B}\rho _{S}^{\alpha }(t)  \notag \\
&&+Tr_{B}L_{SB}\tint\limits_{0}^{t}d\tau e^{L_{0}\tau }L_{SB}\rho _{B}\rho
_{S}^{\alpha }(t-\tau ).  \label{23}
\end{eqnarray}

The obtained Eqs. (\ref{23}) contain (as compared to the standard case) the
additional terms in the kernels governing the evolution of $\rho _{S}(t)$
and $\rho _{S}^{\alpha }(t)$, which account for initial correlations (terms
with $G_{SB}(t,t_{0})$ and $G_{SB}^{t}(t)$).

However, Eqs. (\ref{23}) are still a time-convolution and non-Markovian
ones. To make them time local, we will employ an approach used for obtaining
a time-convolutionless GME (\cite{Shibata1,Shibata2}). It follows from Eqs. (%
\ref{1'''}) and (\ref{9a}), that 
\begin{eqnarray}
F(t-\tau ) &=&U^{-1}(t,t-\tau )F(t)=U^{-1}(t,t-\tau )[f_{r}(t)+f_{i}(t)], 
\notag \\
F_{\alpha }(t-\tau ) &=&e^{-L\tau }F_{\alpha }(t)=e^{-L\tau }[f_{r}^{\alpha
}(t)+f_{i}^{\alpha }(t)].  \label{24}
\end{eqnarray}%
By applying the projector $P$ (\ref{13}) to this relation, using the
definitions (\ref{14}), (\ref{20'}) and that $f_{i}(t)\thicksim \varepsilon $%
, $f_{i}^{\alpha }(t)\thicksim \varepsilon $, we obtain in the zero
approximation in $\varepsilon ,$ \ 
\begin{eqnarray}
\rho _{S}(t-\tau ) &=&U_{S}^{-1}(t,t-\tau )\rho _{S}(t),\rho _{S}^{\alpha
}(t-\tau )=e^{-L_{S}\tau }\rho _{S}^{\alpha }(t)  \notag \\
U_{S}^{-1}(t,t-\tau ) &=&T_{-}\exp [-\tint\limits_{t-\tau }^{t}d\lambda
L_{S}(\lambda )],  \label{25}
\end{eqnarray}%
where we have also used that $Tr_{B}\exp (-L_{B}\tau )\rho _{B}=1$.

Now, remaining in the Born (second in $\varepsilon $) approximation, Eqs. (%
\ref{23}) can be rewritten in the time-local form as%
\begin{eqnarray}
\frac{\partial \rho _{S}(t)}{\partial t} &=&L_{S}(t)\rho
_{S}(t)+Tr_{B}L_{SB}G_{SB}(t,t_{0})\rho _{B}\rho _{S}(t)  \notag \\
&&+Tr_{B}L_{SB}\tint\limits_{0}^{t-t_{0}}d\tau U_{0}(t,t-\tau )L_{SB}\rho
_{B}U_{S}^{-1}(t,t-\tau )\rho _{S}(t),  \notag \\
\frac{\partial \rho _{S}^{\alpha }(t)}{\partial t} &=&L_{S}\rho _{S}^{\alpha
}(t)+Tr_{B}L_{SB}G_{SB}^{t}(t)\rho _{B}\rho _{S}^{\alpha }(t)  \notag \\
&&+Tr_{B}L_{SB}\tint\limits_{0}^{t}d\tau e^{L_{0}\tau }L_{SB}\rho
_{B}e^{-L_{S}\tau }\rho _{S}^{\alpha }(t).  \label{26}
\end{eqnarray}

Equations (\ref{26}) represent the central result of this section. They are
exact in the second approximation in the system-bath interaction and
homogeneous (completely closed) time-local evolution equations for the
system's statistical operator and correlation function, respectively. No
"molecular chaos"-type or Bogoliubov's principle of weakening of initial
correlations approximation has been used. These equations effectively
account for the influence in time of initial correlations on the system's
evolution process via the functions $G_{SB}(t,t_{0})$ or $G_{SB}^{t}(t)$ (%
\ref{21}) defined for an arbitrary initial statistical operator $F(t_{0})$
or $\rho _{tot}(0)$, respectively.

\subsection{An equilibrium Gibbs initial state{}}

Let us suppose, that up to the moment of time $t_{0}$ the total system is in
an equilibrium state with the Gibbs statistical operator but just after $%
t_{0}$ (at $t>t_{0}$) an external (generally time-dependent) force
(described by the Hamiltonian $H_{ext}(t)$) is applied to a system $S$
driving it from an initial state, i.e.,

\begin{eqnarray}
F(t &\leq &t_{0})=\rho _{eq}=Z^{-1}\exp (-\beta
H),H=H_{S}+H_{B}+H_{SB},t\leq t_{0},  \notag \\
\beta &=&1/k_{B}T,Z=Tr_{S+B}\exp (-\beta H),  \notag \\
H(t) &=&H_{S}(t)+H_{B}+H_{SB},H_{S}(t)=H_{S}+H_{ext}(t),t>t_{0}.  \label{27}
\end{eqnarray}%
In this case, the evolution of the system's statistical operator can be
described by Eq. (\ref{26}) for $\rho _{S}(t)$ with the equilibrium initial
statistical operator for the total system $F(t_{0})=Z^{-1}\exp (-\beta H)$.
For this quite realistic initial equilibrium state we can explicitly find
the initial condition function $G_{SB}(t,t_{0})$ (\ref{21}) by making use of
the following exact identity%
\begin{eqnarray}
e^{-\beta H} &=&e^{-\beta H_{0}}-\dint\limits_{0}^{\beta }d\lambda
e^{-\lambda H_{0}}H_{SB}e^{\lambda H}e^{-\beta H},  \notag \\
H_{0} &=&H_{S}+H_{B}.  \label{28}
\end{eqnarray}%
We also select 
\begin{equation}
\rho _{B}=\rho _{B}^{eq}=\frac{e^{-\beta H_{B}}}{Tr_{B}e^{-\beta H_{B}}}
\label{29}
\end{equation}%
in the definition (\ref{13}) for the projection operator.

Then, in the linear approximation in $H_{SB}$, 
\begin{eqnarray}
\rho _{S}(t_{0}) &=&Tr_{B}F(t_{0})=Tr_{B}e^{-\beta H}/Z  \notag \\
&\thickapprox &Tr_{B}[e^{-\beta H_{0}}-\dint\limits_{0}^{\beta }d\lambda
e^{-\lambda H_{0}}H_{SB}e^{\lambda H_{0}}e^{-\beta
H_{0}}]/Tr_{S+B}[e^{-\beta H_{0}}-\dint\limits_{0}^{\beta }d\lambda
e^{-\lambda H_{0}}H_{SB}e^{\lambda H_{0}}e^{-\beta H_{0}}].  \label{30}
\end{eqnarray}%
It is not difficult to see that%
\begin{equation}
Tr_{B}\dint\limits_{0}^{\beta }d\lambda e^{-\lambda H_{0}}H_{SB}e^{\lambda
H_{0}}e^{-\beta H_{0}}=0,  \label{31}
\end{equation}%
and. therefore, in this approximation,%
\begin{equation}
\rho _{S}(t_{0})=e^{-\beta H_{S}}/Tr_{S}e^{-\beta H_{S}}.  \label{32}
\end{equation}%
where we have used the condition (\ref{16a}).

Thus, in the first approximation in $H_{SB}$

\begin{eqnarray}
\lbrack F(t_{0})-\rho _{B}\rho _{S}(t_{0})][\rho _{B}\rho _{S}(t_{0})]^{-1}
&=&e^{-\beta H}[\rho _{B}^{eq}\rho _{S}(t_{0})]^{-1}-1  \notag \\
&=&-\dint\limits_{0}^{\beta }d\lambda e^{-\lambda H_{0}}H_{SB}e^{\lambda
H_{0}},  \label{33}
\end{eqnarray}%
and, finally, the function defining the influence of initial correlations on
the evolution of the system's statistical operator, given by Eq. (\ref{26}),
is 
\begin{equation}
G_{SB}(t,t_{0})=-U_{0}(t,t_{0})\dint\limits_{0}^{\beta }d\lambda e^{-\lambda
H_{0}}H_{SB}e^{\lambda H_{0}}U_{0}^{-1}(t,t_{0}).  \label{34}
\end{equation}

Likewise, selecting $\rho _{tot}(0)=Z^{-1}\exp (-\beta H)$ and $\rho
_{B}=\rho _{B}^{eq}$ (\ref{29}), we obtain for function $G_{SB}^{t}(t)$ (\ref%
{21})%
\begin{equation}
G_{SB}^{t}(t)=-e^{L_{0}t}\dint\limits_{0}^{\beta }d\lambda e^{-\lambda
H_{0}}H_{SB}e^{\lambda H_{0}}e^{-L_{0}t}.  \label{35}
\end{equation}%
This function describes the influence of initial correlation on the
evolution of the equilibrium correlation function (\ref{9a}) (defined with
the $\rho _{tot}(0)=Z^{-1}\exp (-\beta H)$) according to the second Eq. (\ref%
{26}).

It is interesting to note, that the terms describing the influence of
initial equilibrium correlations, given by Eqs. (\ref{34}) and (\ref{35}),
coincide with those obtained in works \cite{Los PHYSA 2018,Los JSP 2017}
(dealing with the polaron mobility) by different method of converting the
inhomogeneous Nakajima-Zwanzig equation into the homogeneous one based on
the identity (\ref{28}), which, however, is applicable only for initial
Gibbs state (\ref{27}) for the total system. The method suggested in this
work (see Sec 2) is applicable for any initial state.

\subsection{A system interacting with the Boson field}

Let us consider the case when a system interacts with the Boson field which
acts as the bath, i.e., we assume that 
\begin{eqnarray}
H_{B} &=&\sum_{k}\hbar \omega _{k}b_{k}^{+}b_{k},  \notag \\
H_{SB} &=&\sum\limits_{k}[C_{k}(S)b_{k}+C_{k}^{+}(S)b_{k}^{+}],  \label{36}
\end{eqnarray}%
where $\hbar \omega _{k}$ is the energy of the field quantum characterized
by the set of quantum numbers $k$, $b_{k}^{+}$, $b_{k}$ are the
Bose-operators of creation and annihilation of the field quantum, and $%
C_{k}(S)$ is an operator acting on a system $S$.

For simplicity, we assume that an external force (contributing to $L_{S}(t)$%
) in Eq. (\ref{26}) for system's statistical operator $\rho _{S}(t)$ is weak
and thus disregard it in the initial correlation and collision terms (linear
response regime). In this case (and putting $t_{0}=0$), the mentioned terms
in Eqs. (\ref{26}) for $\rho _{S}(t)$ and $\rho _{S}^{\alpha }(t)$ become
formally identical.

Then, the term of initial correlations in Eq. (\ref{26}) for $\rho _{S}(t)$
with the use of (\ref{34}), (\ref{35}) (equilibrium initial state), and (\ref%
{36}), can be written as%
\begin{eqnarray}
I_{S}(t,t_{0}) &=&Tr_{B}L_{SB}G_{SB}(t,t_{0})\rho _{B}\rho
_{S}(t)=-PL_{SB}e^{L_{0}t}\int\limits_{0}^{\beta }d\lambda e^{-\lambda
H_{0}}H_{SB}e^{\lambda H_{0}}e^{-L_{0}t}\rho _{B}^{eq}\rho _{S}(t)  \notag \\
&=&i\int\limits_{0}^{\beta }d\lambda \sum\limits_{k}\{e^{-i\omega
_{k}(t-i\lambda )}(1+N_{k})[C_{k}(S),C_{k}^{+}(S_{t,\lambda })\rho _{S}(t)] 
\notag \\
&&+e^{i\omega _{k}(t-i\lambda )}N_{k}[C_{k}^{+}(S),C_{k}(S_{t,\lambda })\rho
_{S}(t)]\},  \label{37}
\end{eqnarray}%
where we put $\hbar =1$ and 
\begin{equation}
C_{k}(S_{t,\lambda })=e^{-iH_{S}t}e^{-\lambda
H_{S}}C_{k}(S)e^{iH_{S}t}e^{\lambda H_{S}},C_{k}^{+}(S_{t,\lambda
})=e^{-iH_{S}t}e^{-\lambda H_{S}}C_{k}^{+}(S)e^{iH_{S}t}e^{\lambda H_{S}}.
\label{38}
\end{equation}%
Here we also used, that for $H_{B}$, given by (\ref{36}),%
\begin{eqnarray}
e^{-iH_{B}t}b_{\mathbf{k}}e^{iH_{B}t} &=&e^{i\omega _{\mathbf{k}}t}b_{%
\mathbf{k}},e^{-iH_{B}t}b_{\mathbf{k}}^{+}e^{iH_{B}t}=e^{-i\omega _{\mathbf{k%
}}t}b_{\mathbf{k}}^{+},  \notag \\
e^{-\lambda H_{B}}b_{k}e^{\lambda H_{B}} &=&e^{\lambda \omega
_{k}}b_{k},e^{-\lambda H_{B}}b_{k}^{+}e^{\lambda H_{B}}=e^{-\lambda \omega
_{k}}b_{k}^{+}  \notag \\
&<&b_{k}b_{k_{1}}>_{B}=0,<b_{k}^{+}b_{k_{1}}^{+}>_{B}=0,  \notag \\
&<&b_{k}b_{k_{1}}^{+}>_{B}=(1+N_{k})\delta
_{kk_{1}},<b_{k}^{+}b_{k_{1}}>_{B}=N_{k}\delta _{kk_{1}},  \notag \\
&<&...>_{B}=Tr_{\Sigma }(...\rho _{B}^{eq}),N_{k}=[\exp (\beta \omega
_{k})-1]^{-1}  \label{39}
\end{eqnarray}

Likewise, we have for the term of Eq. (\ref{26}), which can be identified
with a collision integral, 
\begin{eqnarray}
C_{S}(t,t_{0}) &=&Tr_{B}L_{SB}\tint\limits_{0}^{t}d\tau e^{L_{0}\tau
}L_{SB}\rho _{B}^{eq}e^{-L_{S}\tau }\rho _{S}(t)  \notag \\
&=&-\int\limits_{0}^{t}d\tau \sum\limits_{k}\{e^{-i\omega _{k}\tau
}(1+N_{k})[C_{k}(S),C_{k}^{+}(S_{\tau })\rho _{S}(t)]  \notag \\
&&+e^{i\omega _{k}\tau }N_{k}[C_{k}^{+}(S),C_{k}(S_{\tau })\rho
_{S}(t)]+h.c\}.  \label{40}
\end{eqnarray}%
where 
\begin{equation}
C_{k}(S_{\tau })=e^{-iH_{S}\tau }C_{k}(S)e^{iH_{S}\tau },C_{k}^{+}(S_{\tau
})=e^{-iH_{S}\tau }C_{k}^{+}(S)e^{iH_{S}\tau }.  \label{41}
\end{equation}

The same expressions with the substitution $\rho _{S}(t)\rightarrow \rho
_{S}^{\alpha }(t)$ are valid for the initial correlations and collision
terms of Eq. (\ref{26}) for system's correlation function (see (\ref{20''''}%
)). We note, that in the adopted approximation, the collision integral (\ref%
{40}) is valid for any initial condition (not only for the equilibrium Gibbs
state (\ref{27})).

\section{Quantum oscillator (Fano-like model)}

The results, given by Eqs. (\ref{37}) and (\ref{40}), define the evolution
equations (\ref{26}) for the system's density matrix and correlaton function
with account for initial correlations. These results hold for any system's
Hamiltonian $H_{S}(t)$, any small parameter of the system-Boson field
interaction $C_{k}(S)$, and equilibrium initial state of the total system.

Let us further specify the system and consider\ the following Hamiltonian
for a single-mode cavity system (could be a nanocavity in nanostructures or
photonic crystals) interacting with a Boson reservoir%
\begin{eqnarray}
H(t &=&0)=\omega _{c}a^{+}a+\sum\limits_{k}\omega
_{k}b_{k}^{+}b_{k}+\sum\limits_{k}V_{k}(ab_{k}^{+}+a^{+}b_{k}),  \notag \\
H(t &>&0)=H_{ext}^{S}(t)+H(t=0),  \label{42}
\end{eqnarray}%
where the first term in $H(t=0)$ is the Hamiltonian of the cavity field with
frequency $\omega _{c}$ (corresponds to the system Hamiltonian $H_{S}$), $%
a^{+}$ and $a$ are the creation and annihilation operators of the cavity
field. The second term (corresponds to $H_{B}$) describes the environment as
the reservoir of the infinite Boson modes, and the third term ($H_{SB}$) is
the the system-bath interaction with the coupling strength $V_{k}$. The
Hamiltonian (\ref{42}) represents a Fano-type model of a localized state
coupled with a continuum \cite{Fano}. Thus, we can now calculate explicitly
the initial correlation term (\ref{37}) and the collision term (\ref{40})
for Eqs. (\ref{26}) by putting in these equations 
\begin{eqnarray}
C_{k}(S) &=&V_{k}a^{+},C_{k}^{+}(S)=V_{k}a,  \notag \\
C_{k}(S_{\tau }) &=&e^{-iH_{S}\tau }C_{k}(S)e^{iH_{S}\tau }=V_{k}e^{-i\omega
_{c}t}a^{+},  \notag \\
C_{k}^{+}(S_{\tau }) &=&e^{-iH_{S}\tau }C_{k}^{+}(S)e^{iH_{S}\tau
}=V_{k}e^{i\omega _{c}t}a,  \notag \\
C_{k}(S_{t,\lambda }) &=&e^{-iH_{S}t}e^{-\lambda
H_{S}}C_{k}(S)e^{iH_{S}t}e^{\lambda H_{S}}=V_{k}e^{-i\omega _{c}(t-i\lambda
)}a^{+}  \notag \\
C_{k}^{+}(S_{t,\lambda }) &=&e^{-iH_{S}t}e^{-\lambda
H_{S}}C_{k}^{+}(S)e^{iH_{S}t}e^{\lambda H_{S}}=V_{k}e^{i\omega
_{c}(t-i\lambda )}a.  \label{43}
\end{eqnarray}

For initial correlation term the result is%
\begin{eqnarray}
I_{S}(t) &=&i\int\limits_{0}^{\beta }d\lambda
\sum\limits_{k}V_{k}^{2}\{e^{-i(\omega _{k}-\omega _{c})(t-i\lambda
)}(1+N_{k})[a^{+},a\rho _{S}(t)]  \notag \\
&&+e^{i(\omega _{k}-\omega _{c})(t-i\lambda )}N_{k}[a,a^{+}\rho _{S}(t)]\} 
\notag \\
&=&i\sum\limits_{k}V_{k}^{2}\{e^{-i(\omega _{k}-\omega _{c})t}(1+N_{k})\frac{%
e^{(\omega _{c}-\omega _{k})\beta }-1}{\omega _{c}-\omega _{k}}[a^{+},a\rho
_{S}(t)]  \notag \\
&&+e^{i(\omega _{k}-\omega _{c})t}N_{k}\frac{e^{(\omega _{k}-\omega
_{c})\beta }-1}{\omega _{k}-\omega _{c}}[a,a^{+}\rho _{S}(t)]\},  \label{44}
\end{eqnarray}%
and the collision term is given by%
\begin{eqnarray}
C_{S}(t) &=&-\int\limits_{0}^{t}d\tau \sum\limits_{k}V_{k}^{2}\{e^{-i(\omega
_{k}-\omega _{c})\tau }(1+N_{k})[a^{+},a\rho _{S}(t)]  \notag \\
&&+e^{i(\omega _{k}-\omega _{c})\tau }N_{k}[a,a^{+}\rho _{S}(t)]  \notag \\
&&+e^{i(\omega _{k}-\omega _{c})\tau }(1+N_{k})[\rho _{S}(t)a^{+},a]  \notag
\\
&&e^{-i(\omega _{k}-\omega _{c})\tau }N_{k}[\rho _{S}(t)a,a^{+}]\}  \notag \\
&=&-\int\limits_{0}^{t}d\tau \sum\limits_{k}V_{k}^{2}\{e^{-i(\omega
_{k}-\omega _{c})\tau }(1+N_{k})[a^{+},a\rho _{S}(t)]  \notag \\
&&+e^{i(\omega _{k}-\omega _{c})\tau }N_{k}[a,a^{+}\rho _{S}(t)]+h.c\}
\label{45}
\end{eqnarray}

Thus, we have the following equations for system's statistical operator and
equilibrium correlation function in the case of the Hamiltonian (\ref{42})

\begin{eqnarray}
\frac{\partial \rho _{S}(t)}{\partial t} &=&L_{S}(t)\rho
_{S}(t)+I_{S}(t)+C_{S}(t),  \notag \\
\frac{\partial \rho _{S}^{\alpha }(t)}{\partial t} &=&L_{S}\rho _{S}^{\alpha
}(t)+I_{S}^{\alpha }(t)+C_{S}^{\alpha }(t),  \label{46}
\end{eqnarray}%
where $I_{S}^{\alpha }(t)$ and $C_{S}^{\alpha }(t)$ are defined by Eqs. (\ref%
{44}), (\ref{45}) with the substitution $\rho _{S}(t)\rightarrow \rho
_{S}^{\alpha }(t)$ for the second Eq. (\ref{46}), and $L_{S}(t)$ differs
from $L_{S}$ by the term conditioned \ by an external driving force.

It is instructive to consider Eqs. (\ref{46}) neglecting the initial
condition terms. Then, let us assume that the integration over $\tau $ in (%
\ref{45}) can be extended to infinity, i.e., that we are interested in the
evolution on the relaxation timescale $t\thicksim \tau _{rel}\gg \left\vert
\omega _{k}-\omega _{c}\right\vert ^{-1}$ (the Markov approximation). We
also define the integrals over $\tau $ as%
\begin{equation}
\int\limits_{0}^{\infty }d\tau e^{\pm i(\omega _{k}-\omega _{c})\tau
}=\lim_{\eta \rightarrow +0}\int\limits_{0}^{\infty }d\tau e^{\pm i(\omega
_{k}-\omega _{c})\tau -\eta \tau }=\pi \delta (\omega _{k}-\omega _{c})\pm iP%
\frac{1}{\omega _{k}-\omega _{c}},  \label{46a}
\end{equation}%
where $P$ stands for the integral principal value. As a result, we obtain
the following expression for the collision term%
\begin{eqnarray}
C_{S}(t) &=&-i\Delta \omega _{c}[a^{+}a,\rho _{S}(t)]+J(\omega
_{c})(1+N_{c})[a\rho _{S}(t)a^{+}-\frac{1}{2}\{a^{+}a,\rho _{S}(t)\}]  \notag
\\
&&+J(\omega _{c})N_{c}[a^{+}\rho _{S}(t)a-\frac{1}{2}\{aa^{+},\rho
_{S}(t)\}],t\thicksim \tau _{rel}\gg \left\vert \omega _{k}-\omega
_{c}\right\vert ^{-1},  \notag \\
&&\Delta \omega _{c}=P\int\limits_{0}^{\infty }\frac{d\omega }{2\pi }\frac{%
J(\omega )}{\omega _{c}-\omega },J(\omega )=2\pi
\sum\limits_{k}V_{k}^{2}\delta (\omega _{k}-\omega ),  \notag \\
N_{c} &=&N(\omega _{c}),\{A,B\}=AB+BA.  \label{46b}
\end{eqnarray}

This is the standard Lindblad form for a collision term of a quantum
oscillator (see, e.g. \cite{Breuer}), where $\Delta \omega _{c}$ is a shift
of a frequency $\omega _{c}$ due to interaction with a bath and $J(\omega )$
is the bath spectral density.

Let us now consider Eqs. (\ref{46}) with the contributions of the initial
correlation. We will try to obtain the solution of Eqs. (\ref{46}) for any
timescale and, therefore, will not use the Markov approximation for these
equations (Eq. (\ref{46b}) is only applicable for a large timescale).
Considering, e.g., the equations for $\langle a\rangle _{S}^{t}=Tr_{S}[a\rho
_{S}(t)]$ and $\langle a^{+}\rangle _{S}^{t}=Tr_{S}[a^{+}\rho _{S}(t)]$ and
using (\ref{44}) and (\ref{45}), we easily obtain%
\begin{eqnarray}
\langle aI_{S}(t)\rangle _{S} &=&Tr_{S}[aI_{S}(t)]=i\gamma _{i}(t)\langle
a\rangle _{S}^{t},\gamma _{i}(t)=\sum\limits_{k}V_{k}^{2}e^{-i(\omega
_{k}-\omega _{c})t}(1+N_{k})\frac{e^{(\omega _{c}-\omega _{k})\beta }-1}{%
\omega _{c}-\omega _{k}},  \notag \\
\langle a^{+}I_{S}(t)\rangle _{S} &=&-i\gamma _{i}^{+}(t)\langle
a^{+}\rangle _{S}^{t},\gamma _{i}^{+}(t)=\sum\limits_{k}V_{k}^{2}e^{i(\omega
_{k}-\omega _{c})t}N_{k}\frac{e^{(\omega _{k}-\omega _{c})\beta }-1}{\omega
_{k}-\omega _{c}},  \notag \\
\langle aC_{S}(t)\rangle _{S} &=&-\gamma _{c}(t)\langle a\rangle
_{S}^{t},\gamma _{c}(t)=\int\limits_{0}^{t}d\tau
\sum\limits_{k}V_{k}^{2}e^{-i(\omega _{k}-\omega _{c})\tau },  \notag \\
\langle a^{+}C_{S}(t)\rangle _{S} &=&-\gamma _{c}^{+}(t)\langle a^{+}\rangle
_{S}^{t},\gamma _{c}^{+}(t)=\gamma _{c}^{\ast }(t)=\int\limits_{0}^{t}d\tau
\sum\limits_{k}V_{k}^{2}e^{i(\omega _{k}-\omega _{c})\tau },  \label{47}
\end{eqnarray}

Thus, we have from Eqs. (\ref{46}) with account for initial correlations

\begin{eqnarray}
\frac{\partial \langle a\rangle _{S}^{t}}{\partial t} &=&-i\omega
_{c}\langle a\rangle _{S}^{t}-iTr_{s}\{[a,H_{ext}^{S}(t)]\rho
_{S}(t)]\}+i\gamma _{i}(t)\langle a\rangle _{S}^{t}-\gamma _{c}(t)\langle
a\rangle _{S}^{t},  \notag \\
\frac{\partial \langle a^{+}\rangle _{S}^{t}}{\partial t} &=&i\omega
_{c}\langle a^{+}\rangle _{S}^{t}-iTr_{s}\{[a^{+},H_{ext}^{S}(t)]\rho
_{S}(t)]\}-i\gamma _{i}^{+}(t)\langle a^{+}\rangle _{S}^{t}-\gamma
_{c}^{+}(t)\langle a^{+}\rangle _{S}^{t}.  \label{48}
\end{eqnarray}

In order to consider Eqs. (\ref{48}), we should introduce the external force
in the Hamiltonian (\ref{42}), e.g., as%
\begin{equation}
H_{ext}^{S}(t)=E_{0}\exp (i\omega _{d}t)a+E_{0}\exp (-i\omega _{d}t)a^{+},
\label{48a}
\end{equation}%
where $E_{0}$ is the strength of the external driving field with frequency $%
\omega _{d}$. Then, the inhomogeneous terms in Eqs. (\ref{48}) are%
\begin{eqnarray}
iTr_{s}\{[a,H_{ext}^{S}(t)]\rho _{S}(t)]\} &=&iE_{0}\exp (-i\omega _{d}t), 
\notag \\
iTr_{s}\{[a^{+},H_{ext}^{S}(t)]\rho _{S}(t)]\} &=&-iE_{0}\exp (i\omega
_{d}t).  \label{48b}
\end{eqnarray}

For an equilibrium two-time correlation function $\langle \alpha
_{S}(0)a(t)\rangle $ and $\langle \alpha _{S}(0)a^{+}(t)\rangle $ (see (\ref%
{9a})) with $\rho _{tot}=\rho _{eq}$ (\ref{27}) we have%
\begin{eqnarray}
\langle \alpha _{S}(0)a(t)\rangle _{eq} &=&Tr_{S}\{a[\rho _{S}^{\alpha
}(t)]_{eq}\},\langle \alpha _{S}(0)a^{+}(t)\rangle _{eq}=Tr_{S}\{a^{+}[\rho
_{S}^{\alpha }(t)]_{eq}\},  \notag \\
\lbrack \rho _{S}^{\alpha }(t)]_{eq} &=&Tr_{B}e^{Lt}\rho _{eq}\alpha _{S}(0).
\label{49}
\end{eqnarray}%
In this correlation functions case, Eqs. (\ref{47}) hold with the
substitution ,$\langle a\rangle _{S}^{t}\rightarrow \langle \alpha
_{S}(0)a(t)\rangle ,\langle a^{+}\rangle _{S}^{t}\rightarrow \langle \alpha
_{S}(0)a^{+}(t)\rangle $. Therefore, the equations for these correlation
functions are%
\begin{eqnarray}
\frac{\partial \langle \alpha _{S}(0)a(t)\rangle _{eq}}{\partial t}
&=&\gamma (t)\langle \alpha _{S}(0)a(t)\rangle _{eq},\gamma (t)=-i\omega
_{c}+i\gamma _{i}(t)-\gamma _{c}(t),  \notag \\
\frac{\partial \langle \alpha _{S}(0)a^{+}(t)\rangle _{eq}}{\partial t}
&=&\gamma ^{+}(t)\langle \alpha _{S}(0)a^{+}(t)\rangle _{eq},\gamma
^{+}(t)=i\omega _{c}-i\gamma _{i}^{+}(t)-\gamma _{c}^{+}(t).  \label{50}
\end{eqnarray}%
It is worth noting, that Eqs. (\ref{48}) are inhomogeneous due to the
driving Hamiltonian $H_{ext}^{S}(t)$, whereas Eqs. (\ref{50}) for
correlations functions are the homogeneous ones.

In order to illustrate the influence of initial correlations on the
evolution process, let us consider the more simple homogeneous Eqs. (\ref{50}%
), which describe the dynamics of the cavity field fluctuations. They can be
easily solved and the result is 
\begin{eqnarray}
\langle \alpha _{S}(0)a(t)\rangle _{eq} &=&\exp [\Gamma (t)]\langle \alpha
_{S}(0)a(0)\rangle _{eq},\Gamma (t)=\int\limits_{0}^{t}dt^{\prime }\gamma
(t^{\prime })=-i\omega _{c}t+i\Gamma _{i}(t)-\Gamma _{c}(t),  \notag \\
\langle \alpha _{S}(0)a^{+}(t)\rangle _{eq} &=&\exp [\Gamma ^{+}(t)]\langle
\alpha _{S}(0)a^{+}(0)\rangle _{eq},\Gamma
^{+}(t)=\int\limits_{0}^{t}dt^{\prime }\gamma ^{+}(t^{\prime })=i\omega
_{c}t-i\Gamma _{i}^{+}(t)-\Gamma _{c}^{+}(t),  \notag \\
\Gamma _{i}(t) &=&\int\limits_{0}^{t}dt^{\prime }\gamma _{i}(t^{\prime
}),\Gamma _{i}^{+}(t)=\int\limits_{0}^{t}dt^{\prime }\gamma
_{i}^{+}(t^{\prime }),\Gamma _{c}(t)=\int\limits_{0}^{t}dt^{\prime }\gamma
_{c}(t^{\prime }),\Gamma _{c}^{+}(t)=\int\limits_{0}^{t}dt^{\prime }\gamma
_{c}^{+}(t^{\prime }),  \label{51}
\end{eqnarray}%
where the relaxation functions in Eqs. (\ref{51}) after integration over $%
t^{\prime }$ acquire, as it follows from (\ref{47}) and (\ref{50}), the
following form%
\begin{eqnarray}
\Gamma _{i}(t) &=&i\sum\limits_{k}V_{k}^{2}(1+N_{k})[e^{(\omega _{c}-\omega
_{k})\beta }-1]  \notag \\
&&\times \frac{1-\cos (\omega _{k}-\omega _{c})t+i\sin (\omega _{k}-\omega
_{c})t}{(\omega _{k}-\omega _{c})^{2}},  \notag \\
\Gamma _{i}^{+}(t) &=&i\sum\limits_{k}V_{k}^{2}N_{k}[e^{(\omega _{k}-\omega
_{c})\beta }-1]  \notag \\
&&\times \frac{1-\cos (\omega _{k}-\omega _{c})t-i\sin (\omega _{k}-\omega
_{c})t}{(\omega _{k}-\omega _{c})^{2}},  \notag \\
\Gamma _{c}(t) &=&\sum\limits_{k}V_{k}^{2}[\frac{1-\cos (\omega _{k}-\omega
_{c})t+i\sin (\omega _{k}-\omega _{c})t}{(\omega _{k}-\omega _{c})^{2}}+%
\frac{t}{i(\omega _{k}-\omega _{c})}],  \notag \\
\Gamma _{c}^{+}(t) &=&\sum\limits_{k}V_{k}^{2}[\frac{1-\cos (\omega
_{k}-\omega _{c})t-i\sin (\omega _{k}-\omega _{c})t}{(\omega _{k}-\omega
_{c})^{2}}-\frac{t}{i(\omega _{k}-\omega _{c})}].  \label{53}
\end{eqnarray}%
Using the spectral density of the Bosonic reservoir $J(\omega )$ (\ref{46b}%
), the relaxation functions (\ref{53}) can be rewritten as%
\begin{eqnarray}
\Gamma _{i}(t) &=&\frac{1}{\hbar ^{2}}\frac{i}{2\pi }\int\limits_{-\omega
_{c}}^{\infty }d\omega J(\omega _{c}+\omega )[1+N(\hbar \omega _{c}+\hbar
\omega )](e^{-\hbar \omega \beta }-1)  \notag \\
&&\times \frac{1-\cos (\omega t)+i\sin (\omega t)}{\omega ^{2}},  \notag \\
\Gamma _{i}^{+}(t) &=&\frac{1}{\hbar ^{2}}\frac{i}{2\pi }\int\limits_{-%
\omega _{c}}^{\infty }d\omega J(\omega _{c}+\omega )N(\hbar \omega
_{c}+\hbar \omega )(e^{\hbar \omega \beta }-1)  \notag \\
&&\times \frac{1-\cos (\omega t)-i\sin (\omega t)}{\omega ^{2}},  \notag \\
\Gamma _{c}(t) &=&\frac{1}{\hbar ^{2}}\frac{1}{2\pi }\int\limits_{-\omega
_{c}}^{\infty }d\omega J(\omega _{c}+\omega )[\frac{1-\cos (\omega t)+i\sin
(\omega )t}{\omega ^{2}}+\frac{t}{i\omega }],  \notag \\
\Gamma _{c}^{+}(t) &=&\frac{1}{\hbar ^{2}}\frac{1}{2\pi }\int\limits_{-%
\omega _{c}}^{\infty }d\omega J(\omega _{c}+\omega )[\frac{1-\cos (\omega
t)-i\sin (\omega t)}{\omega ^{2}}-\frac{t}{i\omega }],  \notag \\
N(\hbar \omega ) &=&[\exp (\beta \hbar \omega )-1]^{-1},  \label{54}
\end{eqnarray}%
where we recover the Planck constant.\ 

The formulae (\ref{51}), (\ref{54}) give the exact in the second order
(Born) approximation solution to the system (cavity) equilibrium correlation
functions accounting for initial correlations and valid on any timescale (in
contrast to Eq. (\ref{46b})). The evolution of these correlations functions
is time-reversible (invariant to the $t\rightarrow -t$, $i\rightarrow -i$
replacement).

It is interesting to follow the evolution of the correlation functions with
time in more detail. To this end, we introduce the following timescales%
\begin{equation}
\overline{\omega }t\ll 1,\overline{\omega }t\gg 1,  \label{55}
\end{equation}%
where $\overline{\omega }$ is the characteristic bath frequency, e.g., $%
\hbar \overline{\omega }\thicksim k_{B}T$ (see also the definition of the
relaxation time after Eq. (\ref{46})), or introducing the coherence time $%
t_{coh}$,%
\begin{equation}
t\ll t_{coh},t\gg t_{coh},t_{coh}\thicksim \hbar /k_{B}T  \label{55a}
\end{equation}

Let us first consider the case of the small timescale $\overline{\omega }%
t\ll 1$. In this case, we can approximate $\cos (\omega t)=1-\frac{1}{2}%
(\omega t)^{2}$ and $\sin (\omega t)=\omega t$. As a result we obtain for
the correlation functions (\ref{51}) 
\begin{eqnarray}
\langle \alpha _{S}(0)a(t)\rangle _{eq} &=&\exp [\Gamma _{<}(t)]\langle
\alpha _{S}(0)a(0)\rangle _{eq},\overline{\omega }t\ll 1,  \notag \\
\Gamma _{<}(t) &=&-i(\omega _{c}+\Delta _{i}\omega _{c})t-\gamma t^{2}, 
\notag \\
\Delta _{i}\omega _{c} &=&\frac{1}{\hbar ^{2}}\int\limits_{-\omega
_{c}}^{\infty }\frac{d\omega }{2\pi }\frac{J(\omega _{c}+\omega )}{\omega }%
[1+N(\hbar \omega _{c}+\hbar \omega )](e^{-\hbar \omega \beta }-1),  \notag
\\
\gamma &=&\frac{1}{2\hbar ^{2}}\int\limits_{-\omega _{c}}^{\infty }\frac{%
d\omega }{2\pi }J(\omega _{c}+\omega )\{1+[1+N(\hbar \omega _{c}+\hbar
\omega )](e^{-\hbar \omega \beta }-1)\},  \notag \\
\langle \alpha _{S}(0)a^{+}(t)\rangle _{eq} &=&\exp [\Gamma
_{<}^{+}(t)]\langle \alpha _{S}(0)a^{+}(0)\rangle _{eq},\overline{\omega }%
t\ll 1,  \notag \\
\Gamma _{<}^{+}(t) &=&i[\omega _{c}-(\Delta _{i}\omega _{c})^{+}]t-\gamma
^{+}t^{2},  \notag \\
(\Delta _{i}\omega _{c})^{+} &=&\frac{1}{\hbar ^{2}}\int\limits_{-\omega
_{c}}^{\infty }\frac{d\omega }{2\pi }\frac{J(\omega _{c}+\omega )}{\omega }%
N(\hbar \omega _{c}+\hbar \omega )(e^{\hbar \omega \beta }-1),  \notag \\
\gamma ^{+} &=&\frac{1}{2\hbar ^{2}}\int\limits_{-\omega _{c}}^{\infty }%
\frac{d\omega }{2\pi }J(\omega _{c}+\omega )[1-N(\hbar \omega _{c}+\hbar
\omega )(e^{\hbar \omega \beta }-1)]  \label{56}
\end{eqnarray}%
which shows quadratic in time evolution caused not only by collisions (first
terms in $\gamma $, $\gamma ^{+}$) but also by initial correlations (second
terms). There are also the cavity frequency shifts $\Delta _{i}\omega _{c}$, 
$(\Delta _{i}\omega _{c})^{+}$ conditioned exclusively by initial
correlations. We note, that due to the contribution of initial correlations,
these quantities depend on temperature.

A more interesting is the situation on the kinetic timescale, which we
define as%
\begin{equation}
t\gtrsim \tau _{rel}\gg t_{coh}\thicksim \frac{1}{\overline{\omega }}%
\thicksim \frac{\hbar }{k_{B}T},  \label{57}
\end{equation}%
where the coherence time $t_{coh}$ is defined by (\ref{55a}). The time
hierarchy (\ref{57}) is supposed to be realized in the considered case of a
weak system-environment interaction. On the kinetic timescale we can take a
limit $t\rightarrow \infty $ and use in Eqs. (\ref{54}) the following
representations of $\delta $- function 
\begin{equation}
\lim_{t\rightarrow \infty }\frac{1-\cos (\omega t)}{\omega ^{2}}=\pi \delta
(\omega )\left\vert t\right\vert ,\lim_{t\rightarrow \infty }\frac{\sin
(\omega t)}{\omega }=\pi \delta (\omega )  \label{58}
\end{equation}%
Then, the relaxation functions (\ref{54}) become%
\begin{eqnarray}
\Gamma _{i}(t) &=&\Phi (\beta )=\tau _{rel}^{-1}\hbar \lbrack 1+N(\hbar
\omega _{c})]\beta ,\tau _{rel}^{-1}=\frac{J(\omega _{c})}{2\hbar ^{2}}, 
\notag \\
\Gamma _{i}^{+}(t) &=&\Phi ^{+}(\beta )=\tau _{rel}^{-1}\hbar N(\hbar \omega
_{c})\beta ,  \notag \\
\Gamma _{c}(t) &=&\tau _{rel}^{-1}\left\vert t\right\vert +i\Delta \omega
_{c}t,\Gamma _{c}^{+}(t)=\tau _{rel}^{-1}\left\vert t\right\vert -i\Delta
\omega _{c}t,  \notag \\
\Delta \omega _{c} &=&-\frac{1}{2\pi \hbar ^{2}}P\int\limits_{-\omega
_{c}}^{\infty }d\omega \frac{J(\omega _{c}+\omega )}{\omega },  \notag \\
t &\gg &1/\overline{\omega ,}  \label{59}
\end{eqnarray}%
where we observed that $\int\limits_{-\omega _{c}}^{\infty }d\omega J(\omega
_{c})\frac{\delta (\omega )}{\omega }=\int\limits_{-\infty }^{\infty
}d\omega J(\omega _{c})\frac{\delta (\omega )}{\omega }=0$ and assumed that
the integral over $\omega $ is meant as its principal value. Note, that the
frequency shift here $\Delta \omega _{c}$ coincides with that defined in (%
\ref{46b}) (where $\hbar =1$).

Thus, the time evolution of the system correlation functions on the large
timescale is given by the following expressions (see (\ref{51}))%
\begin{eqnarray}
\langle \alpha _{S}(0)a(t)\rangle _{eq} &=&\exp [i\Phi (\beta )]\exp [-i%
\widetilde{\omega }_{c}t-\tau _{rel}^{-1}\left\vert t\right\vert ]\langle
\alpha _{S}(0)a(0)\rangle _{eq}  \notag \\
\langle \alpha _{S}(0)a^{+}(t)\rangle _{eq} &=&\exp [-i\Phi ^{+}(\beta
)]\exp [i\widetilde{\omega }_{c}t-\tau _{rel}^{-1}\left\vert t\right\vert
]\langle \alpha _{S}(0)a^{+}(0)\rangle _{eq},  \notag \\
\widetilde{\omega }_{c} &=&\omega _{c}+\Delta \omega _{c},t\gg 1/\overline{%
\omega }.  \label{60}
\end{eqnarray}

This is the main result of this section: Equations (\ref{60}) differ from
the standard ones (see, e.g., \cite{Quantum Transport}) by the extra phase
factors $\exp [i\Phi (\beta )]$ and $\exp [-i\Phi ^{+}(\beta )]$ which
emerge due to initial correlations (it is easy to verify, that using the
Lindblad equation (\ref{46b}) in the absence of initial correlations, we can
obtain the result (\ref{60}) with no extra phase factors). These extra
factors can influence the observables, in the considered example associated
with the cavity field fluctuations. The result (\ref{60}) also shows, that
although the initial correlations influence the relaxation with time process
(see (\ref{51}) and (\ref{54})), on the large timescale they cease to do
that ($\Phi (\beta )$ does not depend on time) but can contribute to the
observables. It is also interesting to note, that Eqs. (\ref{60}) are not
time-reversible (although Eqs. (\ref{51}) are) and this is due to initial
correlations. If we disregard initial correlations, Eqs. (\ref{60}) become
time-reversible. Thus, irreversibility on the large (asymptotic) timescale ($%
t\rightarrow \infty $) emerges here due to initial correlations.

Generally, the system correlation function determines the response of the
system on the applied driving force, and, if the initial correlations
survive on the kinetic timescale as in the considered above case, they can
influence the kinetic coefficients. For example, the additional phase factor
(of the type given by (\ref{60})) influences the polaron mobility, as it was
demonstrated earlier in work \cite{Los JSP 2017} treating the dynamics of an
electron interacting with a bath of phonons in the polar crystals.

\section{Summary}

We have presented a novel approach to the dynamics of a system coupled with
a bath. The approach is based on the exact homogeneous (completely closed)
generalized master equations for the relevant parts of a system's
statistical operator and correlation function (Eqs. (\ref{8}) and (\ref{9d}%
)). These equations exactly follow from the inhomogeneous Nakajima-Zwanzig
GME without any "molecular chaos"-type approximations and account for
initial correlations in the kernel governing their evolution in time. \ For
a system in a bath, the obtained equations are equivalent to the homogeneous
equations for a system of interest statistical operator and correlation
function (Eqs. (\ref{20}) and (\ref{20'''})). In the second (Born)
approximation in the system-bath interaction these equations reduce to the
time-local ones (\ref{26}) which describe the evolution of a system
interacting with a bath at any timescale and any initial state of the entire
system (system +bath).

Equations (\ref{26}) are the instruments for investigation of a dynamics of
a specific system in a bath. They have been further specialized for the
important and quite realistic initial Gibbs equilibrium state for the whole
system and the bath as a Boson field. The corresponding initial correlation
and collision terms are given by (\ref{37}) and (\ref{40}).

As an application, the quantum oscillator (cavity mode) interacting with a
Boson field (Fano-like model) has been considered. The equations for an
oscillator statistical operator and correlation function have been obtained
(Eqs. (\ref{46})) with the initial correlations and collision terms given by
Eqs. (\ref{44}) and (\ref{45}). The solution (\ref{51}) for the cavity field
fluctuations (correlation functions), where the generators are defined by
Eqs. (\ref{54}), has been obtained. This solution describes the evolution of
the fluctuations of localized cavity mode at any timescale and shows the
influence of initial correlations on the evolution process. For small times,
the evolution is quadratic in time, while at the large timescale (see Eq. (%
\ref{60})), when the Markov approximation is applicable (actually at $%
t\rightarrow \infty $), the initial correlations cease to influence the
dynamics of the cavity mode fluctuations, but the initial correlations terms
survive and result in the additional phase factor in the correlation
functions, which distinguishes this result from the standard one obtained
from the corresponding Lindblad equation,


\begin{thebibliography}{99}
\bibitem{Bogoliubov} N. N. Bogoliubov, \textit{Problems of Dynamical Theory
in Statistical Physics\ }(Gostekhizdat, Moscow, 1946, in Russian); English
transl.: \textit{Stud. Statist. Mech.\ }\textbf{1} (North-Holland,
Amsterdam, 1962).

\bibitem{van Kampen} N. G. van Kampen, J. Stat. Phys. \textbf{115, }1057
(2004).

\bibitem{Los JSTAT 2024} V. F. Los, J. Stat. Mech. 013107 (2024).

\bibitem{Los JSTAT 2022} V.F. Los, J Stat. Mech. \textbf{1}, 013211 (2022).

\bibitem{Nakajima} S. Nakajima, Progress of Theorerical Physics \textbf{20},
948 (1958).

\bibitem{Zwanzig} R. Zwanzig, J. Chem. Phys. \textbf{33}, 1338 (1960).

\bibitem{Wallace} D. Wallace, Probability and irreversibility in modern
statistical mechanics: Classical and quantum, 2021, arXiv: 2104.11223v1.

\bibitem{Kac} M. Kac, \textit{Probability and Related Topics in Physical
Sciences} (Interscience, New York, 1959).

\bibitem{Breuer} H. -P. Breuer, F. Petruccione, \textit{The Theory of Open
Quantum Systems} (Oxford University Press, New York, 2002).

\bibitem{Lindblad} G. Lindblad, Commun. Math. Phys. \textbf{48}, 119 (1976).

\bibitem{Kossakowski et al} V. Gorini, A. Kossakowski, and E. C. G.
Sudarshan, J. Math. Phys. \textbf{17}, 821 (1976).

\bibitem{Redfield 1957} A. G. Redfield, IBM J. Res. Dev. \textbf{1}, 19
(1957).

\bibitem{Blum 1981} K. Blum, \textit{Density Matrix Theory and Applications}
(Plenum Press, New York, London, 1981).

\bibitem{Phys.Rev.A105032208(2022)} Devashish Tupkary, Abhishek Dhar, Manas
Kulkarni, and Archak Purkayastha, Phys. Rev. A \textbf{105}, 032208 (2022).

\bibitem{Phys.Rev.A83032102(2011)} Hua-Tang Tan and Wei-Min Zhang, Phys.
Rev. A \textbf{83}, 032102 (2011).

\bibitem{Phys.Rev.A100042120(2019)} Gerardo A. Paz-Silva, Michael J. W.
Hall, and Howard M. Wiseman, Phys. Rev. A 100, 042120 (2019).

\bibitem{Phys.Rev.X(2020)} S. Alipour, A. T. Rezakhani, A. P. Babu, K. M\o %
lmer, M. M\"{o}tt\"{o}nen, and T. Ala-Nissila, Phys. Rev. X \textbf{10},
041024 (2020).

\bibitem{Prigogine (1962)} I. Prigogine, \textit{Non-Equilibrium Statistical
Mechanics} (Interscience Publishers, New York, 1962).

\bibitem{Shibata1} F. Shibata, Y. Takahashi, and N. Hashitsume, J. Stat.
Phys. \textbf{17}, 171 (1977).

\bibitem{Shibata2} F. Shibata and T. Arimitsu, J. Phys. Soc. Jap. \textbf{49}%
, 891 (1980).

\bibitem{Los JSP 2017} V. F. Los, J. Stat. Phys. \textbf{168}, 857 (2017).

\bibitem{Los PHYSA 2018} V. F. Los, Physica A \textbf{503}, 476 (2018).

\bibitem{Fano} U. Fano, Phys. Rev. \textbf{124}, 1866 (1961).

\bibitem{Quantum Transport} Chan U Lei and Wei-Min Zhang, arXiv: 1011.4570v2.
\end{thebibliography}
\end{document}